# Graphic

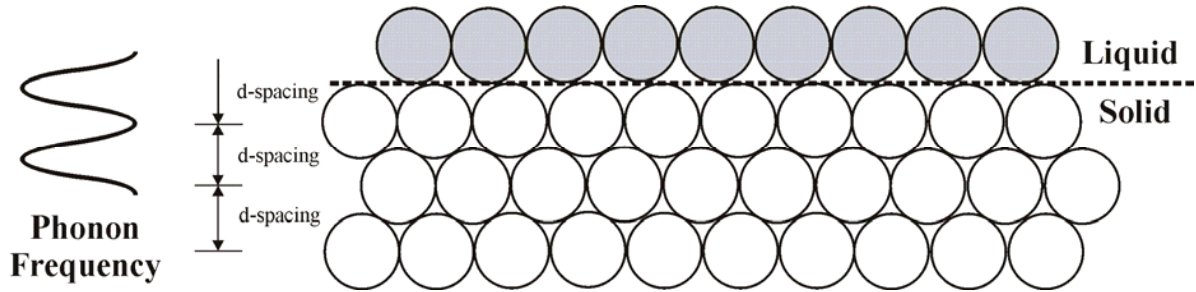

When the wavelength of the phonon is equal or harmonics of the d-spacing then resonance occur and the atomic/molecular sheets are destabilized on the surface of the crystal resulting in melting.

# Melting of crystalline solids

Jozsef Garai

It is suggested that at the melting temperature the thermal phonon vibration is in self-resonance with the lattice vibration of the surface atomic/molecular layer. This self resonance occurs at a well defined temperature and triggers the detachment of the atomic/molecular sheet or platelets from the surface of the crystal. Thermodynamic data of five substances is used to test this hypothesis. The calculated average phonon vibrational wavelengths are equal with or harmonics of the d-spacing of the atomic/molecular sheets. The proposed model is able to explain all of the features of melting.

# 1. INTRODUCTION

Despite a century of research, the mechanism of crystalline-solid melting and the nature of the melt itself still remain something of an enigma [1, 2]. The achievement of a critical level of atomic vibration [3-9], the vanishing of the shear modulus of the crystal [10], and the development of a critical level of defects [11-13] have each been proposed as the primary criterion for the onset of melting. And while they all are important features of melt formation, none of them yield a comprehensive description of the phenomenon.

The distinct characteristics of melting which has to be addressed and explained by any proposed model are

- *vanishing of the shear modulus*

- *existence of the bonds in the liquid* Eventhough atoms/molecules can freely move in the liquid the bonds are not vanished and hold the atoms/molecules together in liquid phase.

- *singularity of the transition* Contrarily to sublimation and vaporization melting of the bulk crystalline solid occurs at well defined pressure and temperature. Thus melting must be some kind of collective atomic/molecular phenomenon which unifies the energies of the atoms/molecules resulting in an abrupt catastrophic collapse of the crystal system.

- *propagation of the melting from the surface* The suggestion melting initiates at the surface and then propagates inward dates back over centuries [14]; however, the microscopic understanding of the phenomenon is still lacking [15]. If the surface is not exposed but instead is coated with a metal layer then crystals do not melt at their normal melting point but rather at a higher one [16].

- *increasing the number of vacancy at melting* One of the proposed explanation for melting is that the lattice instability is induced by a critical level of vacancy. The vacancy concentration on the surface of many metals reaches 10% contrarily to 0.37 % of the bulk [17]. Thus the vacancy induced melting model should be closely tied to surface melting or visa versa.

- *the sort range atomic order in liquids* The melting destroys the long range atomic order of the crystal but sort range order is still present in the liquid. It is well known since the early decades of the last century that liquids has a diffraction halo [18]. This diffraction is attributed to a short range atomic order surrounding any average atom [19].

- *supercooling* Liquids can chill below its freezing point without becoming solid if crystal seeds are absent.

- *nano-size melting* The reduction of the size of the particle results in a melting point depression [20]. The typical reduction of the melting point is linear as a function of the inverse cluster radius [21-23].

- *ultra-fast speed of the phase transition* Femto second electron diffraction study of bismuth showed that the solid-liquid phase transition is extremely fast and completed within 190 fs [24, 25].

Physical model addressing all of these characteristics of melting is developed and tested against experiments of five substances with face centered cubic structure.

## 2. SELF-RESONANCE AT MELTING

It has been suggested that the Debye temperature represents a phonon vibration where the wavelength of the vibration is equivalent with the smallest atomic unit of the crystal structure, which is the unit cell [26]. This suggestion has been tested against the experimental data of monoatomic close-packing arrangements with positive results. Above the Debye temperature the phonon vibration becomes independent and the wavelength of the vibration decreases as the temperature increases. It is suggested that when the phonon vibrational wavelength (or its harmonics) is equal with the d-spacing of the atomic/molecular sheets then self-resonance can occur. This self-resonance can destabilize the surface layer resulting in its detachment. Melt is formed from the detached atomic/molecular layers or platelets.

The self-resonance criterion explains the propagation of the melting from the surface, the singularity of the phase transformation, the increasing number of vacancies at the surface at melting, and the ultra fast speed of the phase transformation. The presence of atomic/molecular platelets in the liquid is consistent with the short range atomic order and with the observed laminar flow of liquids with the floating zone method [27]. The vibrating atomic sheet/platelet model of liquid has been tested by calculating the latent heat of monoatomic solids with positive result [28].

## 3. TESTING THE HYPOTHESIS

The average phonon frequency at the melting temperature corresponds to an average wavelength $[\bar{\lambda}_m]$ which can be calculated [26] as:

$$\bar{\lambda}_m = \frac{h\upsilon_B}{k_B T_m} \qquad (1)$$

where h is the Planck constant, $k_B$ is the Boltzmann constant, $\upsilon_B$ is the bulk seismic velocity and $T_m$ is the melting temperature. The bulk seismic velocity is calculated as:

$$\upsilon_B = \sqrt{\frac{K_T}{\rho}} = \sqrt{\frac{V_m K_T}{N_A M}}.\qquad(2)$$

where $K_T$ is the isothermal bulk modulus, $\rho$ is the density and M is the mass of the atom. In Eq. (2) the isothermal bulk modulus is used instead of the adiabatic. The difference between the two bulk modulus is usually under 1% which is ignored in this study. The effect of temperature on the bulk modulus at 1 bar pressure is given [29] as:

$$K_{0T} = K_o e^{-\int_{T=0}^{T}\alpha_T \delta dT}\qquad(3)$$

where $\delta$ is the Anderson-Grüneisen parameter, $K_o$ is the bulk modulus at zero pressure and temperature, $\alpha_T$ is the volume coefficient of thermal expansion. Assuming that the temperature effect on volume coefficient of thermal expansion is linear it can be written as:

$$K_{0T} = K_o e^{-\int_{T=0}^{T}\alpha_T \delta dT} \cong K_o e^{-(\alpha_o + \alpha_1 T)\delta T}\qquad(4)$$

where $\alpha_o$ is the volume coefficient of thermal expansion at zero pressure and temperature and $\alpha_1$ is a coefficient. Eventhough, the temperature dependence of the coefficient below the Debye temperature is not linear [30], the linear approximation and Eq. (4) can be used for ambient conditions and at higher temperatures [31] because the introduced error is minor. The molar volume of the solid at the melting temperature is calculated by using the EoS of Garai [31], which is given as:

$$V = nV_o^m e^{\frac{-p}{ap+bp^2+K_o}+(\alpha_o+cp)T+\left(1+\frac{cp}{\alpha_o}\right)^f dT^2}\qquad(5)$$

- 5 -

where a is a linear, b is a quadratic term for the pressure dependence of the bulk modulus, c is linear term for the pressure dependence and d and f are parameters describing the temperature dependence of the volume coefficient of thermal expansion. The theoretical explanations for Eq. (5) and the physics of the parameters are discussed in detail [31]. The frequency (f) of a frictionless harmonic oscillator is

$$f = \frac{1}{2\pi}\sqrt{\frac{k}{m}}, \qquad (6)$$

where k is the force constant and m is the mass. If the force constant of the oscillator reduced to half then the frequency is modified as:

$$f_{0.5k} = f_k \sqrt{0.5}. \qquad (7)$$

Assuming that the force constant of the atoms on the surface is half comparing to the bulk and combining Eq. (1) and (7) gives the phonon wavelength at the melting temperature as:

$$\bar{\lambda}_m = \sqrt{0.5}\frac{h\upsilon_B}{k_B T_m}. \qquad (8)$$

The vibrational motion of crystals is very complex and idealized approach is valid only for highly symmetrical atomic arrangement; therefore, substances with face center cubic (fcc) structures are selected for this study.

It is well established that dislocation glides tend to occur parallel with the atomic plane which has the highest planar density [32]. It is assumed that the structure at melting breaks down along its highest planar density. In face centered cubic structure then the distance between the atomic sheets is:

$$d_{111}^{fcc} = \frac{a_{T_m}}{\sqrt{3}}, \qquad (9)$$

where $a_{T_m}$ is the length of the unit cell at the melting temperature. The length of the unit cell is calculated as:

$$a_{T_m} = \sqrt[3]{\frac{n_a V_{T_m}}{N_A}}, \qquad (10)$$

where $n_a$ is the number of atoms in the unit cell and $V_{T_m}$ is the volume at the melting temperature calculated by Eq. (5). Self resonance of the atomic sheets can occur when

$$n\overline{\lambda}_m = d(T_m) \qquad \text{where} \qquad n \in \mathbb{N}^*. \qquad (11)$$

Using the available thermodynamic data of five fcc substances [33-39] the parameters used in Eqs. (4)-(5) have been determined by unrestricted fitting (Tab. 1). Using these parameters the phonon wavelength at the surface [Eq. (8)] and the distance between the vibrating atomic/molecular sheets [Eq. (9)] is calculated at the melting temperature. The results are given in table 2. The calculated $\overline{\lambda}_m / d(T_m)$ ratios indicates that melting occurs when the thermal phonon vibration is firs (Al), second (Au, Pt, MgO, NaCl) harmonics of the d-spacing of the atomic/molecular sheets. The hypothesis, at the melting temperature the thermal phonon vibration is in resonance with the vibration of the atomic/molecular sheets on the surface of the crystal, is conformed for the investigated fcc structures.

## 4. PROPOSED MODEL FOR MELTING

With exception of Al the structure of the crystal breaks down at higher harmonics and not the first one. Why the crystal structure remains stable at the phonon frequency with lower harmonics is investigated.

If atoms are tightly packed, as they are in solid phase, then the interatomic potentials of neighboring atoms will overlap. The maximum of the overlapping potentials surfaces corresponds to a metastable transition state (Fig. 1). The transition state explains the existence of the bonds in liquid phase and the vanishing of the shear modulus.

If resonance occurs below the transition state then the crystal structure remains intact. Thus the melting temperature must be equal or higher than the temperature relating to the metastable transition state. If the energy of the transition state is high then temperatures corresponding to first or may be even higher self-resonance stage might remain inactive.

It is suggested that the atomic/molecular sheets/platelets are detached from the surface when the average energy of the atoms is beyond the transition state and self-resonance occurs. The additional requirement for melting is that energy able to overcome a viscous drag resistance of the liquid must be supplied [28]. This energy is the latent heat. Melting occurs when three criterion are simultaneously satisfied (Fig. 1). One, the average energy of the atoms is higher than the transition state. Two, the phonon vibration is in resonance with the atomic/molecular sheets at the surface. Three, extra energy able to overcome a viscous drag resistance is available.

## 5. SUPERCOOLING

Liquids crystallize at the melting/freezing temperature when a seed crystal or nucleus is present. If this solid interface is missing then the pure liquid phase can be maintained far below the freezing temperature [40].
The suggested explanation for the solidification at the melting/freezing temperature is that the energy taken out from the liquid seizes the translational movement of the atomic platelets which stuck on the surface of solid interface. This solidification process is like a reverse of melting. If solid surface interface is not available then the atomic platelets, despite their movement is seized, stay in the liquid. The energy of the atoms in the platelet is higher than the energy of the transition state; therefore, upon force the platelets can move and the liquid phase preserved (Fig. 2). Theoretically the liquid phase can be maintained down to the temperature corresponding to the transition state.

## 6. NANO-SIZE MELTING

The melting temperature of nano-scale materials are hundreds of degrees lower than the bulk materials [20, 41, 42]. This phenomenon has attracted tremendous interest both theoretically and experimentally for decades [43, 44]. Here I would like to show only that the presented model offers a viable explanation for nano-size melting.

The current consensus on the nano-size particle melting is that the surface atoms are more weakly bond than atoms of the bulk material because the fewer nearest neighbors which results in less constrained thermal motion and lower melting temperature [45]. It is well established that the lattice parameters of the nano-size particles tends to decrease with their size [46-49]. The higher atomic density of the nano-size particles indicates bond strengthening rather then bond weakening of the surface atoms.

The relationship between the stronger bonding and lower melting temperature can be explained by the proposed model. The stronger bonding corresponds to stronger force constant. Thus the difference between bulk and surface layer force constants is smaller than the assumed 2:1 ratio. It can be seen from Eqs. (7) and (8) that the resonance between the atomic/molecular sheet and phonon frequency occurs at lower melting temperature if the force constant of the surface layer is increased. Thus the stronger bonding of the surface layers lowers the melting temperature of nano-size particles because the resonance frequency is reached at lower temperature.

## 7. CONCLUSIONS

It is suggested that the detachment of the atomic/molecular layers from the surface of the crystal is triggered by self-resonance between the thermal phonon and the lattice vibration of the atomic/molecular sheets. Melt is formed from the detached atomic/molecular sheets/platelets. This hypothesis is tested using the thermodynamic data of five substances with face centered cubic crystal structures. The calculated the wavelength of the phonon vibration is first, and second harmonics of the d-spacing of the atomic/molecular sheets.

Incorporating this new element into previous models it is suggested that at melting three criteria have to be simultaneously satisfied. The average energy of the atoms must be higher than the transition state; the phonon vibration must be in resonance with the atomic/molecular sheets at the surface; and extra energy able to overcome a viscous drag resistance must be available. Satisfying these three criteria gives a comprehensive physical description of melting which addresses and explains all of the features of melting.

Although the model should be universally applicable, the structure of the vibrating units is probably more complicated then single or coupled atomic/molecular sheets if the melt is formed from lower symmetry crystals.

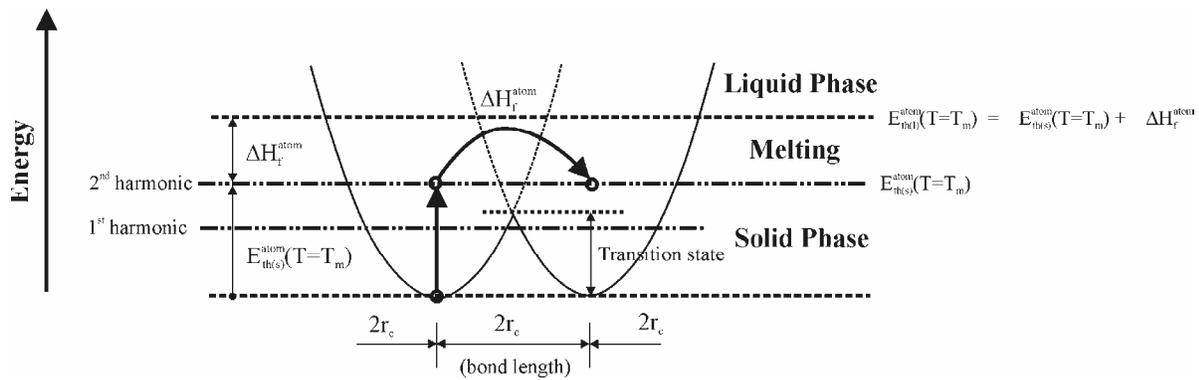

**Fig. 1.** Schematic figure of the energies required for melting. Melting occurs when the average energy of the atoms is higher than the transition state, the thermal phonon vibration is in self-resonance with the lattice vibration of the surface layer and energy required to overcome the viscous drag is available. Self-resonance is effective when the corresponding energy is higher than the transition state.

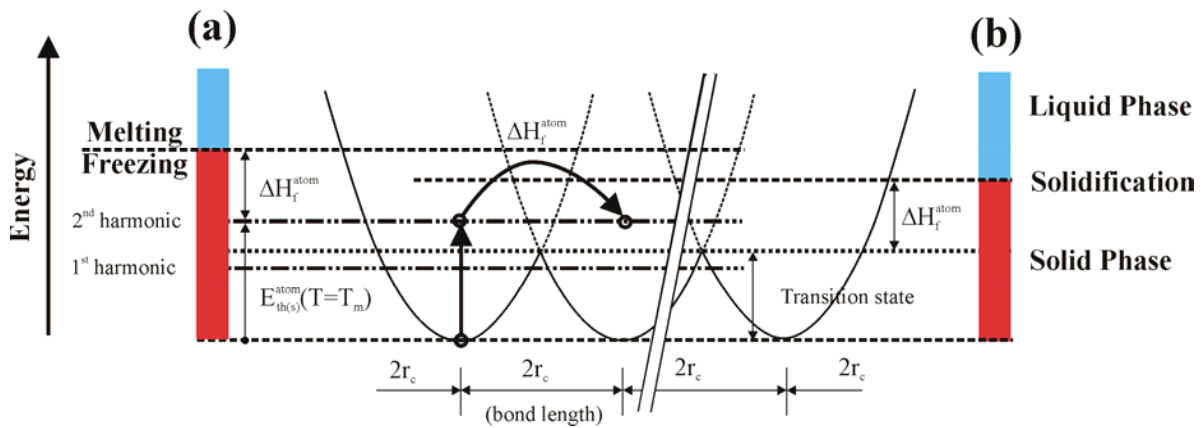

**Fig. 2.** Schematic figure of the energies required for freezing. (a) Solid interface is present and the liquid solidifies at the melting/freezing point. (b) Supercooling occurs when solid interface is lacking. The liquid can be chill down to the temperature representing the transition point.

**Table 1.**

Thermodynamic parameters used for the calculations. The parameters are determined by unconstrained fitting using the data collected from the literature [33-39].

| | Thermodynamic Parameters used for | $V_o$ [cm$^3$] | $K_o$ [GPa] | $K_o'$ | $\alpha_o$ [×10$^{-5}$ K$^{-1}$] | $\alpha_1$ [×10$^{-9}$ K$^{-2}$] | $\delta$ | a | b [×10$^{-3}$] | c [×10$^{-7}$] | d [×10$^{-9}$] | f |
|---|---|---|---|---|---|---|---|---|---|---|---|---|
| Eq. 4 | Al[34, 37] | | 91.00 | 3.769 | 3.284 | 0 | 12.39 | | | | | |
| | Au[34, 35] | | 185.45 | 5.097 | 2.820 | 10.66 | 5.221 | | | | | |
| | Pt[33, 34, 35, 36] | | 254.65 | 6.025 | 1.750 | 5.520 | 4.737 | | | | | |
| | MgO[38 & ref. there in] | | 159.89 | 4.212 | 2.919 | 6.697 | 3.550 | | | | | |
| | NaCl[39] | | 257.35 | 4.901 | 6.451 | 53.85 | 4.127 | | | | | |
| Eq. 5 | Al[34, 37] | 9.804 | 100.00 | | 1.184 | | | 1.983 | -3.145 | -0.820 | 10.24 | 44.4 |
| | Au[34, 35] | 10.119 | 177.16 | | 2.690 | | | 2.458 | -5.704 | -2.492 | 10.09 | 5.4 |
| | Pt[33, 34, 35, 36] | 9.041 | 282.40 | | 1.554 | | | 2.234 | -3.375 | -0.283 | 8.631 | 38.2 |
| | MgO[38 & ref. there in] | 11.149 | 164.95 | | 2.736 | | | 1.717 | -2.126 | -1.502 | 7.893 | 12.0 |
| | NaCl[39] | 26.280 | 287.38 | | 6.573 | | | 1.736 | -7.619 | -20.29 | 70.15 | 13.9 |

**Table 2.**

Calculated average phonon wavelength and d-spacing of the surface layer at the melting temperature.

| Substance | Crystal structure | Bulk Sound Velocity at $T_m$ [ms$^{-1}$] | $T_m$ [K] | Phonon frequency at $T_m$ [$\times 10^{13}$] [s$^{-1}$] | Wavelength of phonon vibration at $T_m$ [Å] | Unit Cell at $T_m$ [Å] | d-spacing at $T_m$ [Å] | Ratio of the wavelength and d-spacing | Harmonics |
|---|---|---|---|---|---|---|---|---|---|
| Al | fcc | 3043.1 | 934 | 1.376 | 2.211 | 4.05 | 2.338 | 0.95 | 1 |
| Au | fcc | 2734.3 | 1337 | 1.970 | 1.388 | 4.14 | 2.390 | 0.58 | 2 |
| Pt | fcc | 3091.9 | 2042 | 3.009 | 1.027 | 4.01 | 2.315 | 0.44 | 2 |
| MgO | fcc | 5546.0 | 2950 | 4.346 | 1.276 | 4.41 | 3.118 | 0.41 | 2 |
| NaCl | fcc | 2798.0 | 1074 | 1.582 | 1.768 | 5.88 | 4.158 | 0.42 | 2 |